\documentclass[a4paper]{article}
\pdfoutput=1

\usepackage[utf8]{inputenc}
\usepackage[T1]{fontenc}
\usepackage[english]{babel}
\usepackage{epigraph}

\PassOptionsToPackage{hyphens}{url}
\usepackage{ifpdf}
\ifpdf
\usepackage[pdftex,breaklinks=true,bookmarks=true,pdfborder={0 0 0}]{hyperref}
\else
\usepackage[breaklinks=true,bookmarks=true]{hyperref}
\fi

\usepackage{listings}
\usepackage{makeidx}
\makeindex

\usepackage{amsfonts}
\usepackage{amsmath}
\usepackage{cleveref}
\usepackage{graphicx}
\usepackage[caption=false]{subfig}

\newcommand{\eg}{e.g.~}
\newcommand{\cf}{cf.~}

\usepackage{enumitem}
\setlist{nolistsep}
\usepackage{multirow}
\usepackage{array}

\begin{document}

\author{R. Poss, S. Altmeyer, M. Thompson, R. Jelier}
\title{“Aca 2.0 Q\&A”\\Usage scenarios and incentive systems\\for a
  distributed academic publication model}

\maketitle

\begin{abstract}
  “Academia 2.0” is a proposal to organize scientific publishing
  around true peer-to-peer distributed dissemination channels and
  eliminate the traditional role of the academic publisher.  This
  model will be first presented at the 2014 workshop on Rproducible
  Research Methodologies and New Publication Models in Computer
  Engineering (TRUST'14) in the form of a high-level overview, so as
  to stimulate discussion and gather feedback on its merits and
  feasibility. This report complements the 6-page introductory article
  presented at TRUST, by detailing the review processes, some use
  scenarios and answering the reviewer's comments in detail.
\end{abstract}

\clearpage

\setcounter{tocdepth}{2}
\tableofcontents

\clearpage

\epigraph{I fail to see how it is our job a scientists to keep
  publishers viable; in hindsight, no one believes we should've
  subsidised horse trainers and blacksmiths to compete with the rise
  of the automobile.}{Merijn Verstraaten}

\section*{Prologue}
\addcontentsline{toc}{section}{Prologue}

The present report is intended to be extended over time, as additional
use cases, issues or opportunities are envisioned. Suggestions for
improvements or new materials are welcome, and the contributor list
can be adjusted accordingly. The latest version of this document, if
any, can be retrieved from \url{http://arxiv.org/abs/1404.7753}.

\section{Introduction}\label{sec:intro}

Supposing the publishing industry did not exist, what could we achieve
to optimize scientific dissemination of knowledge and progress, short of
recreating the publishing industry? This is what the “Academia 2.0”
proposal~\cite{poss.14.trust} addresses.

\begin{figure}
\centering
\fbox{\includegraphics[width=\textwidth]{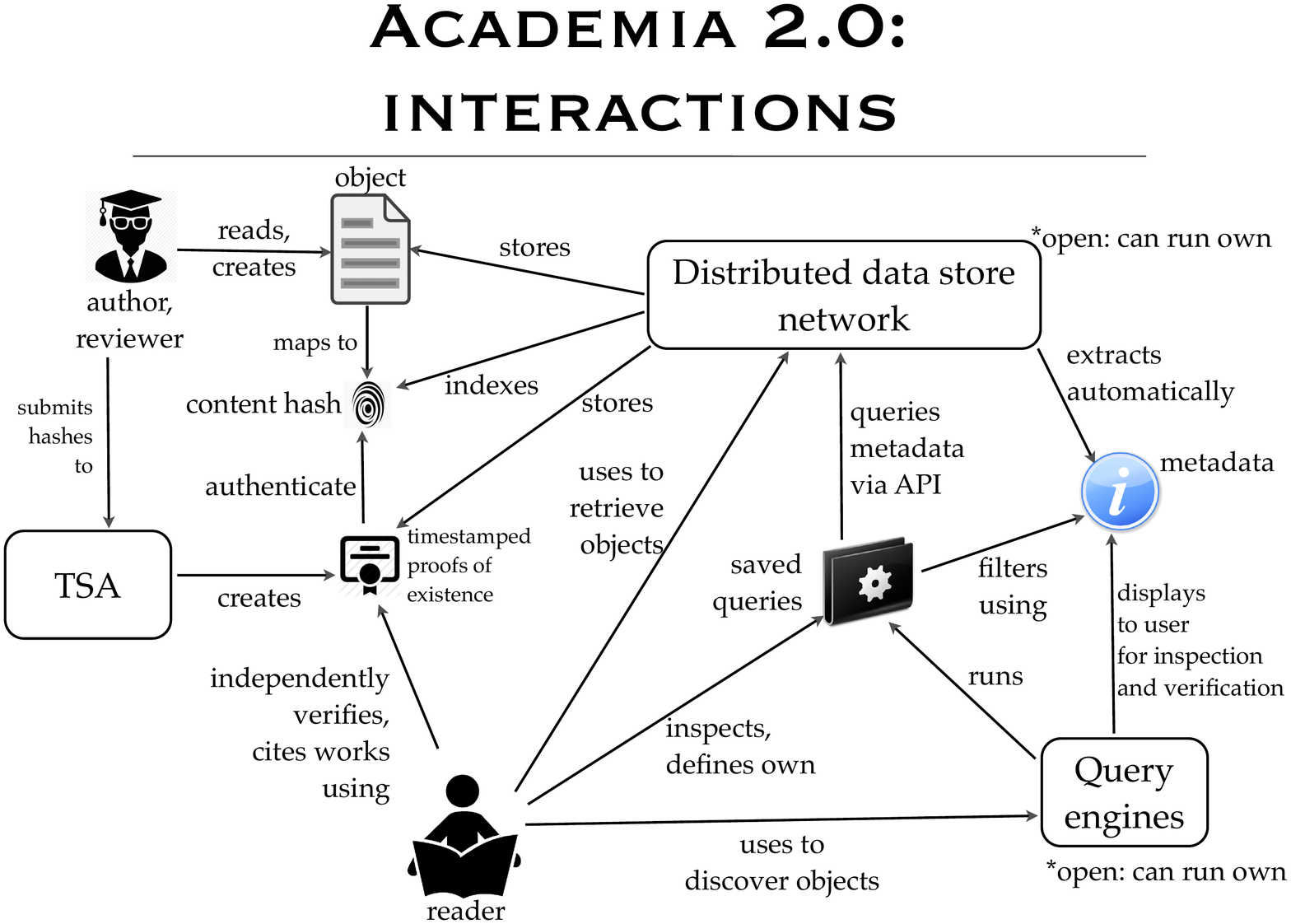}}
\caption{Overview of interactions in Academia 2.0.}\label{fig:overview}
\end{figure}

The proposal can be summarized as follows:
\begin{itemize}
\item every researcher can self-publish online; including reviews of other works whenever they
wish;
\item works are securely timestamped and identified by a fingerprint,
 and optionally title and
author list;
\item public organizations are responsible for publishing
reviews that reviewers wish to keep anonymous while retaining
accountability;
\item a new semantic object, the
``post-hoc citation'' can be used to assert prior work, influence or
plagiarism relationships when they are discovered only after
publication;
\item a new distributed infrastructure serves for document
indexing and lookup; it also provides public and free interfaces for
search and syndication, to aggregate and distribute, on-demand,
relevant documents: per field of expertise, geographical area, social
affinity, or relevance to a topic. See \cref{fig:overview}
\end{itemize}

Following this proposal, a large number of questions have been raised
by the TRUST review committee. The process of answering these questions
also contributes to the elaboration of our model; a synthetic answer
suitable for a summary would fail to convey the more nuanced aspects
of our proposal.

For example, many questions have arisen regarding the organization
of peer review: how to annotate works with measures of their appreciation
by other members of the scientific community. For this, we elaborate
a more complete structured answer in \cref{sec:reviews}.

In particular, the reviewers have highlighted that there have been
already multiple efforts in the past to make the review process of
scientific works more transparent.  Our position is that the “Academia
2.0” proposal is \emph{complementary} to these previous efforts, by
focusing on a different set of goals (\cf the first question in
\cref{sec:qa} and section 2 of the TRUST article). Nevertheless, we
also review the relationship between our work and contemporary efforts
in open peer review below in \cref{sec:rel}.

Similarly, the proposal enables a new incentive system, which in turn
enables new scenarios for the way scientists interact
with each other. We have highlighted two such scenarios in the TRUST
submission; the reviewers have asked for more details and alternate
scenarios, which we provide in \cref{sec:scenarios}.

We then outline architectural aspects from a technical perspective in
\cref{sec:arch}, and address the remaining questions in \cref{sec:qa}.

\section{Peer review}\label{sec:reviews}

The proposal promotes open peer review; we propose to explore
alternative approaches where the review process is transparent, and/or
where the text of reviews is public.  If so desired, the identity of
reviewers may be optionally hidden by an identity escrow service.

The basic ``building block'' is the review object: a document with
semantic fields that identify separately the work(s) being reviewed by
their document handles, the body of the review (text and/or grades),
the author of the review, and which \emph{review process} was used
to produce the review object.

\begin{table}
\caption{Example review object: a TRUST'14 review}\label{tab:exreview}
\scriptsize
\begin{tabular}{|l|p{.85\textwidth}|}
\hline
Author & ``Anonymous reviewer 12 mandated by the TRUST'14 program committee'' \\
\hline
Title  & ``Review of submission 1'' \\
\hline
Target document handle & \{ ``Academia 2.0: removing the publisher middle-man while retaining impact'', [
``Raphael Poss'', ``Sebastian Altmeyer'', ``Mark Thompson'', ``Rob Jelier'' ], ``sha256/e83b0a9861eec4906f52d269056925bd0692c77882ee54d0a62eb876cc61be69'' \} \\
\hline
Overall evaluation & 2/3, higher is better \\
\hline
Reviewer's confidence & 4/5, higher is better \\
\hline
Reviewer comments & The paper presents the idea to have an open publication model. The reviewing, indexing etc will be done in a crowd fashion using semantic technology. The paper is well written and presents novel ideas. My major concern for now is that the value of semantic technology is overestimated. It is also not clear how it can be made sure that not only too few reviews will be available and thus a wrong impression about the work is made visible. \\
\hline
\hline
Review process start date & 2014-03-14 \\
\hline
Review process end date & 2014-04-14 \\
\hline
Review process characteristics & \begin{itemize}
\item Author identity known to reviewer at start of process
\item Review committee known to author before start of process
\item Review object not released publicly before end of process
\item Reviewed object not released publicly before end of process
\end{itemize} \\
\hline
Review process coordinators & \{
 ``Grigori Fursin''  (INRIA, France),
 ``Bruce Childers''  (University of Pittsburgh, USA),
 ``Alex K. Jones''   (University of Pittsburgh, USA),
 ``Daniel Mosse''    (University of Pittsburgh, USA) \} \\
\hline
Reviewer identity escrow & \{
    ``Jose Nelson Amaral'' (University of Alberta, Canada),
    ``Calin Cascaval'' (Qualcomm, USA),
    ``Jack Davidson'' (University of Virginia, USA),
    ``Evelyn Duesterwald'' (IBM, USA),
    ``Lieven Eeckhout'' (Ghent University, Belgium),
    ``Eric Eide'' (University of Utah, USA),
    ``Sebastian Fischmeister'' (University of Waterloo, Canada),
    ``Michael Gerndt'' (TU Munich, Germany),
    ``Christophe Guillon'' (STMicroelectronics, France),
    ``Shriram Krishnamurthi'' (Brown University, USA),
    ``Hugh Leather'' (University of Edinburgh, UK),
    ``Anton Lokhmotov'' (ARM, UK),
    ``Mikel Lujan'' (University of Manchester, UK),
    ``David Padua'' (University of Illinois at Urbana-Champaign, USA),
    ``Christoph Reichenbach'' (Johann-Wolfgang Goethe Universitat Frankfurt, Germany),
    ``Arun Rodrigues'' (Sandia National Laboratories, USA),
    ``Reiji Suda'' (University of Tokyo, Japan),
    ``Sid Touati'' (INRIA, France),
    ``Jesper Larsson Traff'' (Vienna University of Technology, Austria),
    ``Petr Tuma'' (Charles University, Czech Republic),
    ``Jan Vitek'' (Purdue University, USA),
    ``Vladimir Voevodin'' (Moscow State University, Russia),
    ``Vittorio Zaccaria'' (Politecnico di Milano, Italy),
    ``Xiaoyun Zhu'' (VMware,  USA) \} \\
\hline
\end{tabular}
(Note: this example reflects a real review)
\end{table}

An example is given in \cref{tab:exreview}. This review object
represents a possible encoding in our proposal of an actual review
produced by the process of blind review of our submission to the TRUST
workshop.

As the example illustrates, review objects must be self-contained: the
semantic data for evaluation scores embeds information on how to
interpret the scores; the characteristics of the review process are
embedded in the review object; the full list of individuals that serve
as escrow for the identity of the reviewer is given explicitly, for
accountability\footnote{In our example from \cref{tab:exreview}, the
  full program committee is listed as extrow. However, in some
  conferences/journals, only the chairs know the identity of
  reviewers. We believe this is not an issue, because in case of
  investigation the program committee will be able to redirect the
  investigation to the chair as needed.} Although not present in this
example, we suggest also including contact information for the people
involved, and optionally digital signatures to disambiguate homonyms.

Note that this entire review object is itself a publishable document, with a
handle reusing its first two fields and a fingerprint of the entire
object.

\subsection{Characteristics of the review process}\label{sec:chars}

We propose to characterize review processes by the following
properties:

\begin{itemize}
\item the start and end dates of the process, and who is coordinating;
\item the point in time when the author identity is known to the
  reviewer(s). For example: ``at the start of the review process'' (open or simple
  blind reviews), ``at the end of the review process'' (traditional
  double blind review), ``after a first version of the review object
  is released'' (double blind review, PLDI-style);
\item how the reviewer identity is known to the public. This can be
  either ``open'' (the reviewer publishes the review object under
  their own name), or ``anonymized'': the author field only lists a
  number or pseudonym, and an identity escrow (\eg the program
  committee) is accountable for relating the review to an actual
  reviewer upon investigation if necessary (\cf \cref{sec:anon}
  below);
\item when the reviewer identity is known to the public. For example:
  ``prior to the start of the review process'' (eg. conferences, where
  the program committee is published in the call for papers), ``at the
  end of the review process'' (eg. journals, where reviewers are
  selected only after works are submitted for review);
\item when the review text is published. For example ``immediately''
  (arbitrarily soon after the reviewer has finished populating the
  review object) or ``at the end of the review process'' (reviews held
  by a review committee until all reviewers have contributed, to
  guarantee reviewer independence).
\end{itemize}

``Academia 2.0'' does not mandate a specific review process: whether the
review is blind, double blind, open or otherwise. Instead, each review
object should embed meta-information about the process that was used
to produce it. Our example in \cref{tab:exreview} illustrates use
of a blind review process, although it is not named as such.
The benefit of this \emph{declarative} approach are threefold:
\begin{itemize}
\item it decouples the review object from the specific platform (in
  this case, Easychair for the TRUST workshop) that was used to
  coordinate the process, thereby ensuring the process can be
  understood and checked long after the specific platform has disappeared;
\item it enables a more diverse set of review processes, while
  ensuring full transparency;
\item the users of the review object can make their own opinion about
  the side effects of the review process (whether multiple reviews on
  the same work were independent, etc).
\end{itemize}

Naturally, this characterization is backward-compatible with the
review processes already used in the existing ecosystem. We illustrate
this in \cref{tab:rchars}.

\begin{table}
\caption{Characteristics of review processes}\label{tab:rchars}
\scriptsize
\hspace{-10em}\begin{tabular}{|p{.2\textwidth}|p{.16\textwidth}|p{.125\textwidth}|p{.17\textwidth}|p{.13\textwidth}|p{.18\textwidth}|p{.32\textwidth}|}
\hline
\multirow{2}{.2\textwidth}{Author identity known to reviewer}
           & \multicolumn{2}{p{.25\textwidth}|}{%
              Reviewer identity known}
                                     & \multicolumn{2}{l|}{%
                                       Review text availability} & \multirow{2}{.18\textwidth}{Reviewed work public}  & \multirow{2}{*}{Found in...} \\
\cline{2-5}
           & How        & When       & When       & To whom      &                       &      \\
\hline \hline
Prior      & Open       & Immediate  & Immediate  & Public       & Prior                 & Forums, blogs, \newline survey articles, \newline ``related work'' sections \\
\hline
Prior      & Anonymized & Prior      & Afterwards & Authors, PC  & Afterwards, only beyond threshold & Traditional blind review\newline (conference) \\
\hline
Prior      & Anonymized,
             to authors only
                        & Afterwards & Afterwards & Authors, PC  & Afterwards, only beyond threshold & Traditional blind review\newline (journal) \\
\hline
Afterwards & Anonymized & Prior      & Afterwards & Authors, PC  & Afterwards, only beyond threshold & Traditional double blind review (conference) \\
\hline
Afterwards & Anonymized,
             to authors only
                        & Afterwards & Afterwards & Authors, PC  & Afterwards, only beyond threshold & Traditional double blind review (journal) \\
\hline
\multicolumn{7}{l}{} \\
\hline
\bf Prior or afterwards & \bf Open or anonymized & \bf Prior or afterwards & \bf Immediate or afterwards & \bf \underline{Public} & \bf Prior or afterwards & \bf Academia 2.0 \\
\hline
\multicolumn{1}{l}{Examples:} & \multicolumn{6}{l}{} \\
\hline
Prior & Open & Prior & Immediate or afterwards & '' & Prior & Author-requested review round on previous publication by author-selected reviewers \\
\hline
Prior & Open & Afterwards & Immediate & '' & Prior & Spontaneous (reviewer-directed) review of an existing publication  \\
\hline
Prior & Anonymized & Afterwards & Afterwards  & '' & Prior & Blind review round on existing publication by committee-selected reviewers \\
\hline
Afterwards & Open or anonymized & Afterwards & Afterwards & '' & Afterwards & Double blind review round on new work by committee-selected reviewers \\
\hline
\end{tabular}
\end{table}

\subsection{Anonymous reviews}\label{sec:anon}

Showing the public identity of reviewers may impact the
integrity of researchers: one may not be able to objectively and
publicly criticize poor work performed by colleagues, or a potential
future colleague in a hiring position.

To compensate for this, we suggest that public organizations
(eg. conference program committees, libraries) propose \emph{review
  anonymization} as an optional service to their local review
contributors. Using this service, the reviews are public but published
under a pseudonym, and the escrow is responsible for keeping track of
real identities. This escrow service for researcher identities would
also protect accountability: consistently poor reviews could then be
tracked to their real authors after suitably authorized
investigations.

\paragraph{How to deal with problematic reviews by anonymous reviewers?}

This would happen as follows: say, the public observes that a significant
number of reviews coming from a single identity escrow are
consistently poor. What to do about this?

At a first level, our model already enables the community to publish
meta-reviews (reviews of the reviews) that denounce their quality,
give them a poor score, and thus influence transitively the ranking of
the articles promoted/discredited by the poor reviews.

We also propose that escrow services provide an investigation
board: if a petition of (external) researchers demands an
investigation regarding a group of reviews, the escrow's investigation
board should \emph{internally} track the reviews to the anonymized
review author (whom they know, as they are escrowing this person's identity)
and request either the retraction of the reviews (by publication of
counter-reviews by the same author) or a clarification/extension of
the reviews. At no point is the anonymous author's identity disclosed
to the external petitioners. If the escrow service is failing to
respond to investigation requests, search queries could subsequently
dismiss \emph{any} review coming from the same escrow.

\subsection{Organization of the review process}\label{sec:rorg}

Open reviews are relatively simple in that they do not require any coordination:
either an author contacts (a) reviewer(s) directly, or reviewer(s) decide
spontaneously to perform a review, and review objects are published as soon as they are
authored with all identities public.

Additional care is required for blind reviews, or review groups where
multiple reviews are requested while providing some confidence the
resulting reviews were authored independently from each other.

\index{next-generation journal}

For conferences and next-generation journals (\cf \cref{sec:newgen}), this coordination could
be performed as usual by the program committee, with the optional
help of a technical review platform like Easychair.

Yet this coordination can also be self-organized directly by a group
of authors and reviewers, using an \textsl{ad hoc} ``review
committee'' selected for the occasion. For example, a self-coordinated
blind review round could happen as follows:

\begin{itemize}
\item a group of authors prepares a work for publication, and generates
  document handles for them;
\item the authors contact a group of experts in their field, to invite
  them for a review round with pre-determined start and end dates for
  the process;
\item the authors mandate the experts who accept the invitation to, in
  turn, invite candidate reviewers (or decide to become reviewers
  themselves) but without revealing the reviewer's identities;
\item the authors provide privately a copy of their work to the
  experts, together with a template for the review object containing
  the process meta-data; the experts then circulate the work and
  template (privately) to their selected reviewers;
\item during the process (before the end date), the reviewers prepare
  reviews, and generate document handles for them;
\item at the end of the review process, the reviewers release their
  review objects publicly and the authors release their work publicly.
\end{itemize}

Note that beyond the selection of reviewers, the experts selected by
the authors need not be involved in the review process at all. The
coordination of the process does not either require a centralized
process management platform. In particular, the \emph{review platforms
  become ephemeral}: one can deactivate/destroy the specific technical
environment where reviews have been produced, without losing the
reviews, their impact in query engines, and transparency about the
review process.

\subsection{Double blind reviews}\label{sec:dbr}

Double blind reviews is an evaluation process where the authors submit
their work in a way that cannot be traced back to them; and where the
authors do not know who the reviewers are. After the review period,
the identities of authors and reviewers for those works that have been
``accepted'' (evaluated beyond a predetermined threshold) may be
released publicly.

\subsubsection{Organization}

It is possible in ``Academia 2.0'' to organize double-blind reviews, by
extending the process from \cref{sec:rorg} as follows:

\begin{itemize}
\item a group of authors prepares a work for publication that includes their
  names in the author list, and generates a
  fingerprint and certificate of existence X for this non-anonymized version;
\item the authors also prepare an anonymized version of the work;
\item the authors organize the review round as in \cref{sec:rorg},
  with the following adjustments: 1) the intermediate experts should
  not play the roles of reviewers themselves 2) the anonymized version
  of the work is provided for review 3) the review template includes
  the fingerprint and certificate of existence X of the non-anonymized version;
\item during the process (before the end date), the reviewers prepare
  reviews, including the fingerprint X, and generate document handles for them;
\item at the end of the review process, the reviewers release their
  review objects publicly and the authors release the non-anonymized work publicly.
\end{itemize}

\begin{figure}
\centering
\fbox{\includegraphics[width=\textwidth]{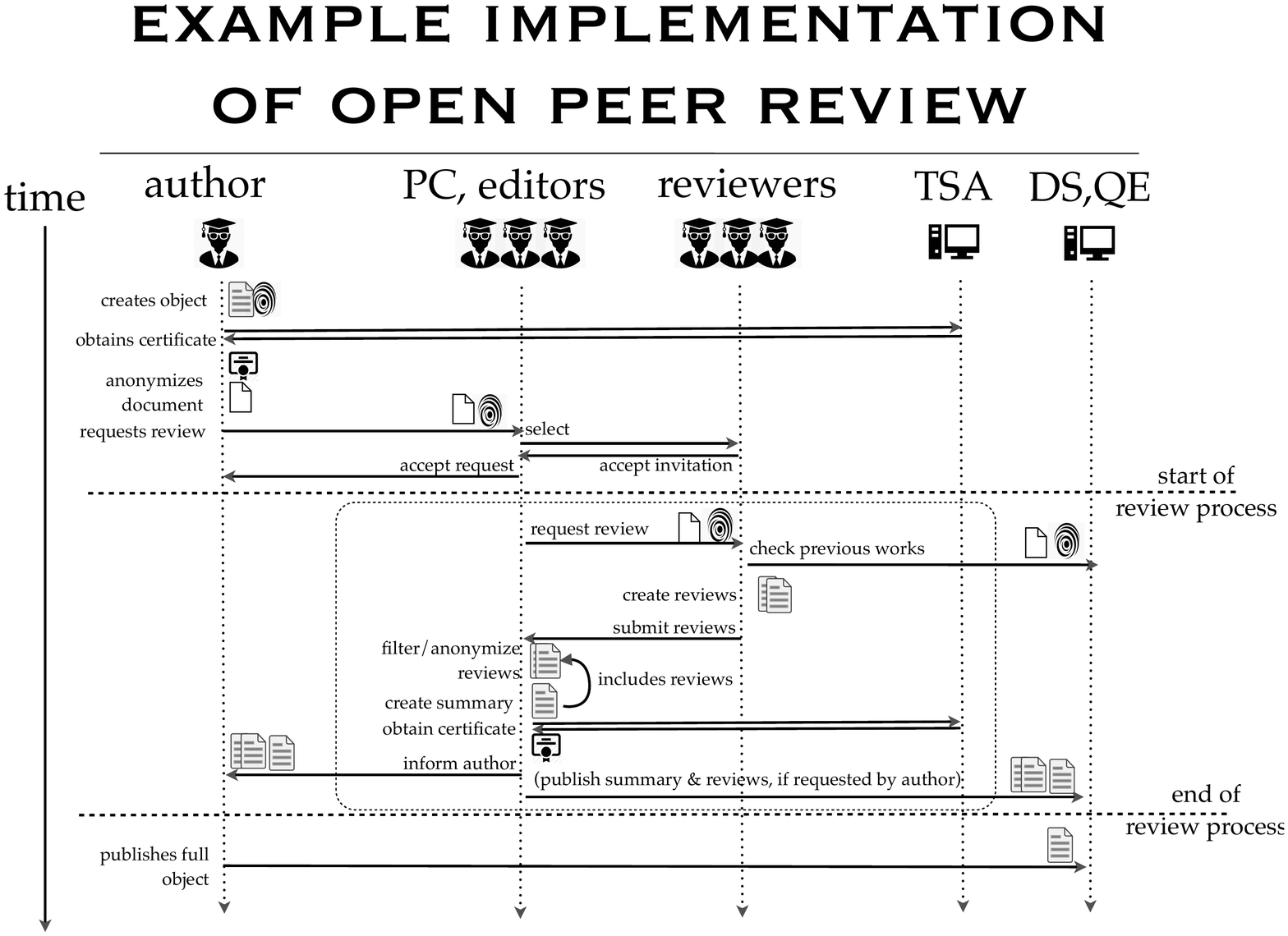}}
\caption{Construction of document handles.}\label{fig:reviews}
\end{figure}

An example implementation is given in \cref{fig:reviews}.

Using the fingerprint of the non-anonymized version in the review
object ensures that the review is securely linked to the
work at the end of the review process, even though the reviewers do
not know the author identities at the time of review.

\subsubsection{Desirability of double blind peer review}

Even though ``Academia 2.0'' enables double blind peer review,
controversy exists in the scientific community as to whether it is at
all desirable.  Although this discussion falls slightly outside of the
scope of this technical report, we include here an argument by
D. Lemire~\cite{lemire.11.dbpr} which summarizes the controversy:

\begin{quote}
How well does it work in practice? You would expect double-blind
reviewing to favor people from outside academia. Yet Blank
reported~\cite{blank.91} that the opposite is true: authors from
outside academia have a lower acceptance rate under double-blind peer
review. Moreover, Blank indicates that double-blind peer review is
overall harsher. This is not a surprise: It is easier to pull the
trigger when the enemy wears a mask.

Meanwhile, there is at best a slight increase in the quality of the
papers due to double-blind peer review~\cite{devries.09}—everything
else being equal. However, not everything is equal under double-blind
peer review. What is the subtext? That somehow, the research paper is
a standalone artifact, an anonymous, standardized piece of LEGO. That
it should not be viewed as part of a stream of papers produced by an
author. It sends a signal that an original research program is a bad
idea. Researchers should be interchangeable. And to assess them, we
might as well count the number of their papers—since these papers are
standard artifacts anyhow.

But that is counter-productive! Research papers are often only
interesting when put in a greater context. It is only when you align a
series of papers, often from the same authors, that you start seeing a
story develop. Or not. Sometimes you only realize how poor someone’s
work is by collecting their papers and noticing that nothing much is
happening: just more of the same.

Researchers must make verifiable statements, but they should also try
to be original and interesting. They should also be going
somewhere. Research papers are not collection of facts, they represent
a particular (hopefully correct) point of view. A researcher’s point
of view should evolve, and how it does is interesting. Yet it is a lot
easier to understand a point of view when you are allowed to know
openly who the authors are.

Are there cliques and biases in science? Absolutely. But the best way
to limit the biases is transparency, not more secrecy. Let the world
know who rejected which paper and for what reasons.
\end{quote}

Remarkably, ``Academia 2.0'' avoids a pitfall
identified by M. Wilson~\cite{wilson.12}, reported by D. Lemire:

\begin{quote}
What if you pick up good ideas from double-blind papers that are later
rejected and remain unpublished? How do you acknowledge the
contribution of the authors of the unpublished work?
\end{quote}

With our proposal, works reviewed by a double-blind process are still
published even if the review scores are unfavorable. Post-hoc
citations can then complement this by asserting influence relationships
later on.

\subsection{Previous work on open peer review}\label{sec:rel}

\begin{table}
\caption{Articles on open peer review published in 2011-2012 in JCN.}\label{tab:pub18}
\begin{tabular}{llp{.6\textwidth}}
Ref. & Author(s) & Title \\
\hline
\cite{bachmann.11.fcn} &
T. Bachmann &
Fair and open evaluation may call for temporarily hidden authorship,
caution when counting the votes, and transparency of the full
pre-publication procedure
\\
\cite{birukou.11.fcn} &
A. Birukou et al. &
Alternatives to peer review: novel approaches for research evaluation
\\
\cite{florian.12.fcn} &
R.V. Florian &
Aggregating post-publication peer reviews and ratings
\\
\cite{ghosh.12.fcn} &
S.S. Ghosh et al. &
Learning from open source software projects to improve scientific review
\\
\cite{hartshorne.12.fcn} &
J.K. Hartshorne et al. &
Tracking replicability as a method of post-publication open evaluation
\\
\cite{hunter.12.fcn} &
J. Hunter &
Post-publication Peer Review:  Opening up Scientific Conversation
\\
\cite{iettogillies.12.fcn} &
G. Ietto-Gillies &
The evaluation of research papers in the XXI century. The Open Peer
Discussion system of the World Economics Association
\\
\cite{kravitz.11.fcn} &
D.J. Kravitz et al. &
Toward a new model of scientific publishing: discussion and a proposal
\\
\cite{kreiman.11.fcn} &
G. Kreiman et al. &
Nine criteria for a measure of scientific output
\\
\cite{kriegeskorte.12.fcn2} &
N. Kriegeskorte &
Open evaluation (OE): A vision for entirely transparent
post-publication peer review and rating for science.
\\
\cite{lee.12.fcn} &
C. Lee &
Open Peer Review by a Selected-Papers Network
\\
\cite{poeschl.12.fcn} &
U. P\"oschl &
Multi-stage open peer review: scientific evaluation integrating the
strengths of traditional peer review with the virtues of transparency
and self-regulation
\\
\cite{priem.12.fcn} &
J. Priem et al. &
Decoupling the scholarly journal
\\
\cite{sandewall.12.fcn} &
E. Sandewall &
Maintaining Live Discussion in Two-Stage Open Peer Review
\\
\cite{walther.12.fcn} &
A. Walther et al. &
{FOSE}: A framework for open science evaluation
\\
\cite{wicherts.12.fcn} &
J.M Wicherts et al. &
Letting the daylight in: reviewing the reviewers and other ways to
maximize transparency in science
\\
\cite{zimmermann.12.fcn} &
J. Zimmermann et al. &
Network-based statistics for a community driven transparent
publication process
\\
\end{tabular}
\end{table}

In 2011-2012, the journal Frontiers in Computational Neurosciences (FCN) ran
a call to collect ideas and evaluations for alternate approaches
to peer review in science. The resulting series of 18
articles on the topic, listed in \cref{tab:pub18}, was then linked together
by one overview from Kriegeskorte et al.~\cite{kriegeskorte.12.fcn}.

The ``emerging consensus''~\cite{kriegeskorte.12.fcn} is a synthetic collection
of \emph{desirable features} that should be available in all new
developments, and \emph{feature choices} that will characterize
different approaches from each other.

The features deemed universally desirable are:

\begin{itemize}
\item full transparency of the evaluation process;
\item evaluative information is semantically encoded, eg. via priority scores;
\item any group or individual can define a formula for prioritizing
  papers, fostering a plurality of evaluative perspectives;
\item the evaluation process includes at least written reviews and
  numerical ratings;
\item reviews and ratings are meta-evaluated (reviews of reviews);
\item paper evaluations are weighted according to both the scientific
  and reviewing performance of participating scientists;
\item the open evaluation process is perpetually ongoing;
\item the new system can evolve from the current one.
\end{itemize}

\textbf{Our proposed model matches all these expressed requirements.}
Note also that our model does not mandate a specific modality for the
review process, and is therefore inclusive of all the proposals linked
to by~\cite{kriegeskorte.12.fcn}. Different modalities could be
exploited in different groups/places over time depending on each
community's local requirements.

Some additional features are proposed in the article group but not deemed universally
desirable by all authors in the group:

\begin{itemize}
\item whether the reviews are digitally signed;
\item whether the author identities are securely authenticated;
\item whether formal statistical inference is a primary feature of the
  evaluation process.
\end{itemize}

\textbf{Our proposal is agnostic to these requirements.} Beyond the
basic certificate of existence provided by our proposed secure
timestamping, published objects can but do not have to include digital
signatures for authors. As to which facilities are provided by query
engines, our proposal welcomes implementations that are based on formal specifications.

Finally, the following feature choices appear polarizing in the group:

\begin{itemize}
\item whether evaluation begins with a closed, pre-publication stage,
  or whether all results are published continuously and \emph{filtered}
  a posteriori;
\item whether queries rank papers according to usage, social web
  information and citations.
\end{itemize}

Here again \textbf{our proposal is agnostic}. A community of
researchers may decide to self-organize and only publish works in our
proposed data stores after they have received a predefined number of
positive reviews from their community. But this is not mandatory. As
to how queries are organized, our proposal distinctly specifies that
any user of a query engine can specify their own queries, and we thus
expect that a diversity of criteria will be exploited in practice. At
any rate, each individual review object will contain meta-data as to
which review process was used (\cf \cref{sec:chars}), so that its characteristics stay
visible over time and can be analyzed later on.

\section{Scenarios and incentives}\label{sec:scenarios}

\subsection{Fear of credit loss}

{\small
(This scenario is already included in the TRUST overview article~\cite{poss.14.trust})
}

\noindent
Some researchers have strong feelings against the mandatory
publication of software, tools and dataset next to research results,
arguing that this openness and transparency will enable
competitor researchers to exploit said tools and then publish results
earlier than the original author. When this happens, the first
researcher has to bear the main cost of the research (time, effort) but
cannot reap the profits (a lot of emphasis in reward systems is put on
which researcher is first to publish), which is quite unfair
indeed. This incentive to avoid publication hinges on both the reality of
unscrupulous researchers, and the reality of the long publication time
for journal articles (commonly up to one or two years between
submission and final acceptance).

Obviously, our proposal eradicates the second factor as publication
then becomes essentially instantaneous. As for unscrupulous
researchers, our proposal has a number of built-in features which are
relevant.

First, certificates of existence can be obtained by a
researcher on his or her preliminary results, \emph{before they are
actually available publicly}: our proposed time authorities can
issue a certificate based only on the fingerprint of a document's content, so
the actual content can remain private until a later date. This
enables a researcher to operate using the following workflow:
\begin{enumerate}
\item prepare the tools;
\item compute some preliminary / proof-of-concept results using them;
\item privately obtain a timestamp certificate, which attests the
  work's existence although it is not published yet;
\item publish the tools;
\item later, when more results are obtained, publish the
  high-level outcomes of the research.
\end{enumerate}
Using this process, if a competitor exploits the
published materials from step \#4 and claims original work similar to
\#3, the first researcher can assert his or a her prior work \textsl{a
  posteriori} using the timestamp certificate.

Moreover, supposing the original researcher did not realize that
further results have been published by competitors without attribution
or failed to request timestamp certificates early, the post-hoc
citation mechanism can be used as well: the research or even a peer
can publish a statement of influence, from the ``real'' original work
to the competitor work, after both have been published and
\emph{regardless of publication order}. In practice:
\begin{enumerate}
\item some work is discovered to be likely ``heavily inspired'' on
  some other work;
\item one or more researchers publish post-hoc citations declaring the
  same;
\item the social network of the interested parties scrutinizes the
  relationship and, when deemed relevant, strengthen the weight of
  the post-hoc citations in the search network by adding
  \emph{positive reviews} for the post-hoc citations themselves.
\end{enumerate}
Using these steps, both original and derived work become linked by the
post-hoc citations. Agreement by peers strengthen the post-hoc
citations. To fully exploit this opportunity, we envision
that query engines list both sides of post-hoc citations when a search
query would otherwise only return one side, together with the relevant
context (including comments) around post-hoc citations.

\subsection{Early publication}

What motivates a contributor to (self-)publish their work early? What
are reasons this might be undesirable?

\paragraph{Pros:}
faster dissemination \& more reuse. This is especially relevant in
technology fields where know-how can become obsolete within 1-2
years. Factoring both communication, evaluation and learning times, a
publication pipeline longer than 1-2 months may make reuse less
attractive than redesigning from scratch.

\paragraph{Cons:}
\begin{itemize}
\item
the earlier a work is published, the less scrutiny it receives
\textsl{prior} to publication. This may result in an average quality
decrease.  \textbf{Solution:} \emph{continuous evaluation}
post-publication, and adequate \emph{post-publication filtering} based on the
current network of reviews and reuse links.
\item each individual publication from an author/group may reflect a
  smaller incremental step since a previous publication. This may
  result in increasing volume of literature for the same research
  output. More volume in turn makes dissemination more difficult
  (other researchers have a limited ``reading
  bandwidth''). \textbf{Solution:} develop an authoring culture where
  \emph{works become distributed}: each publication could be split in
  a 1-2 pages \emph{overview article} with motivation and conclusions,
  with \emph{links to separate published objects} for results,
  methods, tools and analysis. If this happens, the throughput of
  overview articles may increase, (which is desirable), but without
  overloading the reading time of research peers too much.
\end{itemize}

\subsection{File drawer effect and publication bias}

\index{file drawer effect}
\index{publication bias}

What motivates a contributor to avoid (self-)publishing negative results?
How to promote/ease/favor the public availability of negative results?
How to prevent publishers from refusing to publish negative results?

\paragraph{Proposed approach:} ideally, any work on an experiment could be packaged
as publishable objects for intent, methodology, and result sets over
time.  The distributed nature of ``Academia 2.0'' implies that
researchers can obtain secure future-proof identifiers for any object
virtually for free (a certificate of existence only requires a digital
signature of the fingerprint, not the content
itself), so the gathering of identifiers could be fully automated in
experimental workflows, and the negative datasets can be kept stored
at the location where they were initially produced.  Negative results
can then be cited and reused from the same researchers or their peers.

This workflow avoids distribution costs (negative datasets left at
site of production), reviewing costs (no external party involved), and
registration costs (certificate of existence based merely on fingerprint),
while ensuring that negative results are available for reuse shortly
after they are produced.

\subsection{Next-generation journals}\label{sec:newgen}
\index{next-generation journal}

In the current ecosystem, journals provide multiple services:
\begin{enumerate}
\item they centralize document storage;
\item they organize dissemination of documents (publicity and access);
\item they coordinate the review process;
\item they provide an entry point for research communities to quickly identify ``top publications'' in their field.
\end{enumerate}

Journal publishers currently charge high fees, claiming these fees are
needed to cover the costs of services \#1-\#3, and these fees are tolerated
by the research community because they are currently the most
efficient provider of service \#4.

As explained previously~\cite{poss.14.trust}, ``Academia 2.0'' mainly
intends to replace publishers for services \#1-\#3, and drive down the
associated costs.  The proposal also suggests implementing service \#4
using saved queries and syndication, thereby replacing journal publishers entirely.

However, note that \emph{removing the role of publisher does not imply
  that the concept of journal disappears}.  Instead, ``Academia 2.0''
enables \textbf{next-generation journals}, with lower operating costs and
new revenue streams.

\subsubsection{Organization of next-generation journals}

A next-generation journal would be identified by:

\begin{itemize}
\item a group of owners, replacing the role of ``steering committee'' or ``editor'';
\item a site location on the Internet, under control of the owners, which advertises itself as ``a journal'';
\item a set of portal web pages on that site that automatically display the results of searches from pre-saved queries;
\item the definition of the pre-saved queries, created/configured by the owners to automatically select the
  published objects that match the journal's criteria; \emph{these must be kept public} so that users can assess the journal's criteria and
  determine how it chooses to influences the scientific field;
\item \emph{added-value services} that the owners can charge for.
\end{itemize}

The process to create such a journal is simple, and can be thus instantiated at near-zero cost
by any scholarly organization for more fluid competition:

\begin{enumerate}
\item a group of experts self-organizes and appoints itself as a new journal owner;
\item they set up a query engine using the standard ``Academia 2.0'' software on their own hosting location;
\item they define their own pre-saved queries according to their desired criteria;
\item they set up a web site / portal to present query results;
\item if desired, they set up extra added-value (chargeable) services.
\end{enumerate}

From then on, the journal can run ``hands off'', optionally entirely independently from the review process(es) (\cf \cref{sec:reviews}).

\subsubsection{Revenue stream}

This model of next-generation journals prevents journal owners from charging for
the mere distribution and access to the works, since the queries are public
and can thus be arbitrarily reproduced by third parties.

Instead, revenue can be extracted using added value services, for example:
\begin{itemize}
\item guaranteed low-latency access to all linked documents/objects (a document cache);
\item moderated online discussion fora for (group of) selected works;
\item match-making services, to put researchers in contact to other individuals with similar interests in their field;
\item opt-in advertisement services to the press or other organizations.
\end{itemize}

\section{Matters of architecture and implementation}\label{sec:arch}

In short, ``Academia 2.0'' proposes to combine:
\begin{itemize}
\item a new \emph{culture} for scholars, which encompasses open peer review and decentralized storage and filtering;
\item a new \emph{incentive system} which makes low-cost global and long-term access to knowledge possible, while enabling new business forms;
\item a set of \emph{feature requirements} for platforms that are compatible with the ``Academia 2.0'' vision;
\item a federation of \emph{platforms} that satisfy the rquirements and provide the model's components: time stamping, data stores and query engines.
\end{itemize}

\subsection{Current status}

Regarding culture, the growing discontentment of scholars with the existing publishing
ecosystem, and the numerous initiatives to develop open access and
open review, was already pushing the ``Academia 2.0'' agenda before
this proposal was written. The present proposal complements this
cultural shift by connecting it with concrete proposals for a
tool-workflow integration and specific requirements to satisfy the
simultaneous goal of guaranteed cheap and long-term access.

The incentive system is being developed iteratively; some first steps
are visible already in \cref{sec:reviews,sec:scenarios} in this
document.

The rest of this \cref{sec:arch} outlines the feature requirements.

As to platforms, the ``Academia 2.0'' proposal does not mandate a
specific implementation. It is likely that we (the proposal's authors) will
propose an example implementation, for the sake of
illustration; meanwhile, we also acknowledge multiple projects around the world that
develop tools already compatible with the ``Academia 2.0'' vision, including in particular:
\begin{itemize}
\item the Public Knowledge Project (PKP, \url{https://pkp.sfu.ca/}) in Canada, which produces open
  source software platforms for online journals, monograph
  publication, conference organization (incl. publication of
  proceedings) and metadata indexing and search;
\item Cornell University Library's arXiv (\url{http://arxiv.org/}), which provides an online
  service for document registration (time stamping) and long-term
  storage.
\end{itemize}

The PKP's ``Harvester system'', for example, could be seen as a
suitable implementation of an Academia 2.0 data store and query
engine, able to import documents from any OAI-compliant archive. The arXiv repository
can be combined with an index of fingerprints to become a suitable Academia 2.0 data store.

Additional platforms dedicated to organizing conferences or online
journals, for example those provided by the PKP, are
\emph{complementary} to the ``Academia 2.0'' vision and could be
extended to connect to its data stores and query engines.

\subsection{Certificates of existence}

\index{time stamps}
\index{certificate of existence}

The distributed data stores must annotate incoming documents with
certificates of existence (CoEs), that attest they existed at the time
they were submitted. We are interested in time stamping schemes that
can be verified by third parties long after they were generated, and
that do not require to trust one arbitrary party.

We envision two general avenues to define/generate CoEs:
\begin{itemize}
\item \emph{third parties registries}: existing and future registries
 like DOI, the ACM digital library, arXiv.org, etc. can continue
 to have this role; the identifiers they deliver can be reused
 as CoEs in our proposal;
\item \emph{distributed time stemps}: new protocols and systems
  can be devised to generate secure CoEs without the need to trust
  specific third parties.
\end{itemize}

Academia 2.0 promotes the recognition of multiple CoE sources
side-by-side, so as to remove the incentive for a single party to
charge predatory prices for the registration service. Our technical
proposal in \cite{poss.14.ads} follows this requirement.

\paragraph{Overview of existing technologies:}

There exist multiple open standards and technologies to do this,
including RFC 3161 \& 4998, ANSI X9.95 and ISO 18014 (see
list below). Each standard defines how to \emph{package} and
distribute time stamps, and supports multiple different mechanisms to
\emph{generate} them, acknowleging that different applications have
different security goals.

For example, with PKI-based generation mechanisms the authentication is dependent
on the trust user place on the owners of private keys to use them
properly and keep them secret.

Another example is linking-based generation mechanisms. There a
combined (``linked'') hash is published, at fixed time intervals, of
the documents created since the last publication. This is done for
example via newspapers. The authentication is then dependent on the
trust users place on the wide distribution of a newspaper and archival
of previous issues.

\paragraph{Related academic references:}
\begin{itemize}
\item \cite{haber.91} Haber et al. ``How to time-stamp a
  digital document'', 1991.
\item \cite{blibech.06} Blibech et al. ``A New Timestamping Scheme
  Based on Skip Lists'', 2006.
\item \cite{buchmann.09} Buchmann et al. ``Hash-based
  Digital Signature Schemes'', 2009.
\end{itemize}

\paragraph{Current open standards and specifications:}
\begin{itemize}
\item \href{https://tools.ietf.org/html/rfc3161}{RFC 3161}, ``Internet
  X.509 Public Key Infrastructure Time-Stamp Protocol
  (TSP)''. Abstract: \emph{This document describes the format of a
    request sent to a Time Stamping Authority (TSA) and of the
    response that is returned.  It also establishes several
    security-relevant requirements for TSA operation, with regards to
    processing requests to generate responses.}
\item \href{https://tools.ietf.org/html/rfc3628}{RFC 3628}, ``Policy
  Requirements for Time-Stamping Authorities (TSAs)''. Abstract:
  \emph{This document defines requirements for a baseline time-stamp
    policy for Time-Stamping Authorities (TSAs) issuing time-stamp
    tokens, supported by public key certificates, with an accuracy of
    one second or better.}
\item \href{http://webstore.ansi.org/RecordDetail.aspx?sku=ANSI+X9.95-2012}{ANSI
  X9.95},
  ``Trusted Time Stamp Management and Security''. Abstract: \emph{This
  standard specifies the minimum security requirements for the
  effective use of time stamps in a financial services environment.}
\item \href{https://tools.ietf.org/html/rfc4998}{RFC 4998}, ``Evidence
  record syntax''. Abstract: \emph{In many scenarios, users must be
    able prove [sic] the existence and integrity of data, including
    digitally signed data, in a common and reproducible way over a
    long and possibly undetermined period of time.  This document
    specifies the syntax and processing of an Evidence Record, a
    structure designed to support long-term non- repudiation of
    existence of data.}
\item \href{http://www.iso.org/iso/search.htm?qt=18014&type=simple&published=on}{ISO/IEC
  18014},
  ``Information technology -- Security techniques -- Time-stamping
  services''. Abstract: \emph{ISO/IEC 18014 specifies time-stamping
    techniques. It consists of three parts, which include the general
    notion, models for a time-stamping service, data structures, and
    protocols.}
\item \href{http://www.openksi.org/}{OpenKSI}, ``Open working group
  for keyless signature infrastructure''. Abstract: \emph{Keyless
    Signature Infrastructure (KSI) is a web-scale digital
    signature/timestamp system for electronic data. The term ‘keyless’
    denotes that the signatures can be reliably verified without
    assuming secrecy of cryptographic keys. Keyless signatures are not
    vulnerable to key compromise implying that their lifetime is based
    only on the security properties of the cryptographic hash function
    and also that their security properties will survive intact when
    practical quantum computing becomes a reality. }
\end{itemize}

\subsection{Document handles}

\index{document handle}

``Academia 2.0'' is strongly defined by content-based addressing:
works and documents are identified by their content; not by where they
are stored (file name, database record identifier), nor where they
were published (publication name, date, page number). This enables
long-term persistence of documents, robust to changes in organization
names, technology evolution, economical situations, etc.

\begin{figure}
\centering
\fbox{\includegraphics[width=\textwidth]{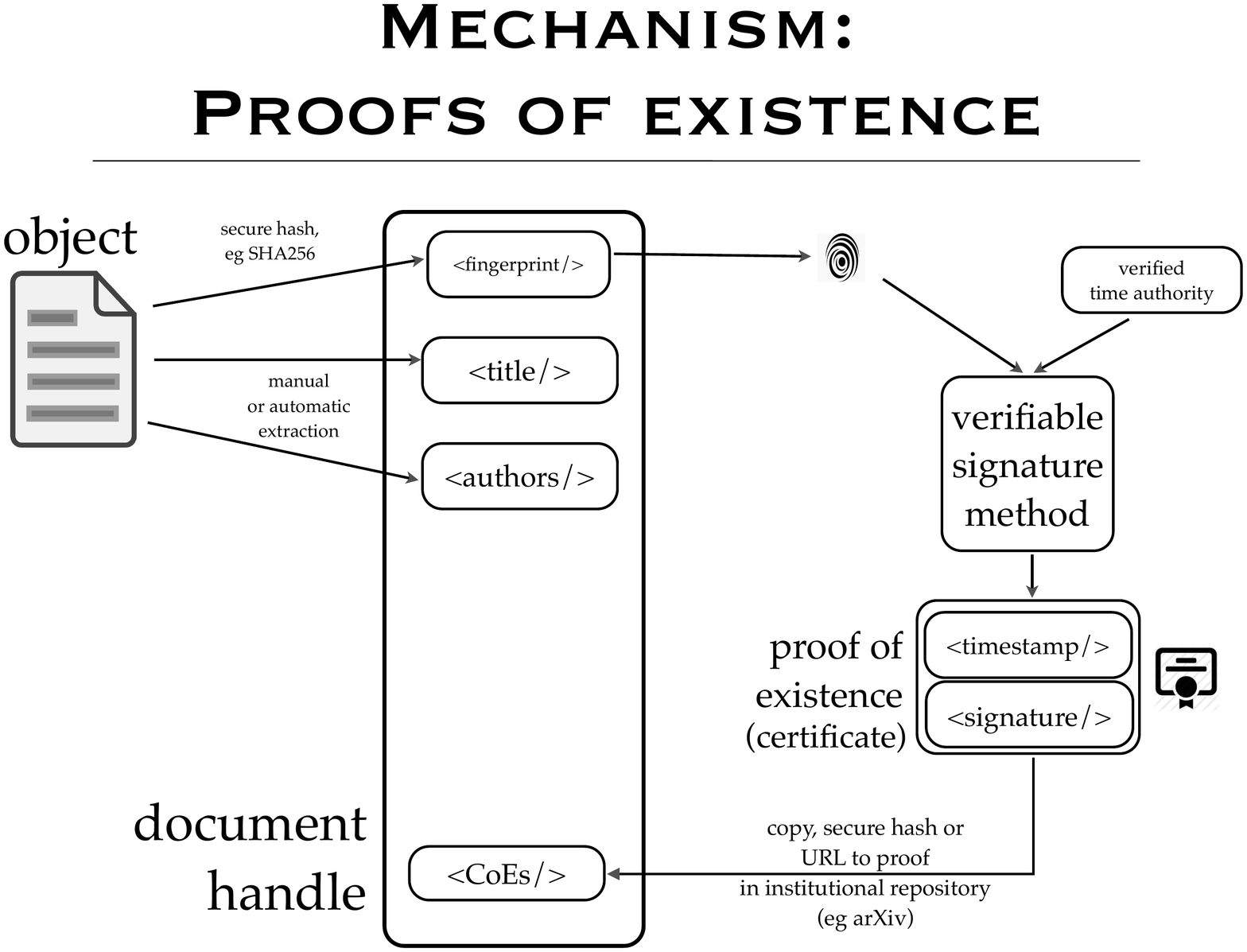}}
\caption{Construction of document handles.}\label{fig:dh}
\end{figure}

The most prominent feature of ``Academia 2.0'' which advertises
content-based addressing to users is the \emph{document handle}: a
4-tuple consisting of the work title, author list, fingerprint and
CoE. See \cref{fig:dh}. The ``fingerprint'' part of a document handle is also the
key that refers to the data of the work identified by the handle:
\emph{data stores} accept request for such a key, and return a copy of
the data as response.  The CoE is a certificate of existence of the data
linked to by the fingerprint, not of the document handle itself.

Note that only the fingerprint and CoE should ultimately be
authoritative: the work's title and author list should be present in
the document itself (and thus captured by the fingerprint and CoE), and
the information only copied in the handle to ease indexing by data
stores. When possible, an implementation can even extract the title
and author list on demand from the document's contents. This makes it
also possible to keep document handles short when the title spans
multiple lines (or sub-titles are present) or the author list is
particularly long (in some fields, scientific articles have hundreds
of authors).

For example, a document handle for version 2 of the present report
could be:

\begin{tabular}{|l|l|}
\hline
Field & Value \\
\hline
Title & Aca 2.0 Q\&A \\
\hline
Authors & "R. Poss", "S. Altmeyer", "M. Thompson", "R. Jelier" \\
\hline
Fingerprint & \texttt{fp:FvYPWVbnhezNY5vdtqyyef0wpvj149A7SquozxdVe3jigg} \\
\hline
CoEs & \texttt{arxiv:2014-05-10:1404.7753v2} \\
\hline
\end{tabular}

In this handle:
\begin{itemize}
\item the fingerprint identifies the PDF document (\cf next section,
  and also~\cite{poss.14.ads});
\item the title field contains a reduced version of the full title
of the document, which further extends to ``Usage scenarios and
incentive systems for a distributed academic publication model'';
\item the author list omits the full first name of each author;
\item the timestamp uses arXiv's authority to assert that the document
  was visible on May 10th, 2014, with arXiv's document identifier as
  signature.
\end{itemize}

\subsection{Object model}

With ``Academia 2.0'' we wish to promote structured data, so as to
support also collections of items (\eg data sets) or academic articles
with separate data streams for print (\eg PDF) and web display (\eg HTML).

We thus propose to develop an \emph{object model} which
supports both singleton files and dictionaries.  Our technical proposal
in~\cite{poss.14.ads} follows this suggestion.

\subsection{Object storage and distribution}\label{sec:storage}

The high-level view of ``Academia 2.0'' is that data stores are
responsible for the persistence of objects over time, maintain
metadata and reply to object queries from users and query engines.

To realize this, we need to reconcile two apparently conflicting
requirements. On the one hand, users value the authority and
seriousness of official institutions such as university libraries, and will
trust them more to persist data in the long term. On the other hand,
users wish unrestricted access to content over the entire network, but
official institutions will likely regulate access in accordance to
affiliation and local licensing rules.

\subsubsection{Suggestion: a dual network}

\index{network!institutional}
\index{network!peer-to-peer}

Our proposed approach is to set up different networks of data stores
simultaneoulsy, with different objectives and policies but using a
common interface (API) to users and query engines. We propose to begin
with the following combination:
\begin{itemize}
\item an \emph{institutional network} of data stores, where each node is managed by a public organization.  This network may:
  \begin{itemize}
  \item restrict the point of submission from authors by affiliation or
  geographical proximity,
  \item maintain explicit and contractual peering agreements between
    data stores,
  \item attribute one or more ``home'' data store to each object, in
    order to attribute the responsibility for long-term storage of an
    object to the organization(s) hosting its ``home'' data store(s);
  \item delete ``old'' copies of objects stored in other stores than their home locations;
\item propagate unsatisfied object requests from users or query engines along peering links until either the
  object is found or a strong confidence is ascertained that the object does
  not exist in the entire network.
\item propose ``quality control'' on submissions, or license-based
  access control to non-public objects.
\end{itemize}
\item a \emph{peer-to-peer network} of data stores, where each node is managed by an individual person. This network may:
\begin{itemize}
\item refuse external submissions for new objects altogether (only each node owner can insert objects locally, for example their own documents);
\item maintain loose peer-to-peer connections with other data stores;
\item implement object queries using a distributed hash table (DHT) protocol;
\item keep transient cached copies of objects gathered by queries, so they become candidates to further searches.
\end{itemize}
\end{itemize}

In short, we propose that ``Academia 2.0'' relies simultaneously on a
network of conventional document repositories with institutional
control over object distribution, and a Bittorrent-like peer-to-peer
file sharing network where distribution is under personal
responsibility of its node owners.

The institutional network would be fast to answer
queries, more certain about unsatisfied queries (``no answer = the
object does not exist''), guarantee long-term storage institutionally,
but may limit free access to objects; whereas the peer-to-peer network would be
slower to answer queries, less certain about unsatisfied queries (``no
answer = the object may or may not exist''), dilute/distribute
responsibility for long-term storage, but will guarantee free access
to objects.

\subsubsection{Side benefits of the dual approach}

\index{network!peer-to-peer}

Our proposal to use a dual design was also intended to solve a major
challenge of ``Academia 2.0'': how to deploy the new model
side-by-side with the existing publishing ecosystem.

In particular, to promote the use of our query engines, it is
relatively important that data stores index and generate metadata for
a large corpus of historical published works, \emph{including those
  held exclusively by existing publishers}. In other words: since
existing publishers disallow redistribution of their content outside
of their distribution channels, how can their works become visible in
Academia 2.0?

Certainly, a proposal to include publisher-protected articles in the
institutional network would imply that the participating institutions
pay a licensing fee to the publishers. The related financial,
logistical and political issues would likely create friction to a
speedy migration to the new model.

This is where the peer-to-peer network would ease the transition, by
exploiting a particular feature of the legal system around academic
publishing: both authors and customers that have already paid once to
receive a copy of a work are allowed to distribute copies to an
audience of students or other peers that contact them directly. In
other words, publishers have already long agreed that direct
peer-to-peer distribution of scholarly content is acceptable as long
as the source of documents is properly linked to an author or licensed
user.

This is a major difference between the rights given by publishers to
scholarly authors and licensed users, and the rights given by film and
music publishers to artists and consumers: neither artists or
consumers receive a license for private redistribution. This is why
peer-to-peer redistribution of works in Academia 2.0 can be built-in
from the get-go, without concern for licensing conflicts.

In practice, the scenario would be covered as follows:
\begin{itemize}
\item both the institutional and peer-to-peer network would ``know''
  about previous documents, identified by their fingerprint; both
  would import metadata from either personal repositories from
  scholars, or from their library collections;
\item however, the institutional network would only know about the existence of the documents, and not store a copy of the objects identified by the metadata;
\item a user wishing to consult a previously published work would then have a choice between:
\begin{itemize}
\item either contacting the legacy publisher directly to obtain a
  license for the content; or
\item query the peer-to-peer network for the fingerprint and receive
  a copy of the content directly from a source willing to share it.
\end{itemize}
\end{itemize}

To ease the transition, it may be possible to identify nodes in the
peer-to-peer network by the identity of their owners, so as to
establish a clear audit trail for document copies. This full
transparency of content propagation will probably make the transition
smoother for the existing publishers.

\section{Questions and Answers}\label{sec:qa}

\noindent
\textbf{Q: What exactly are the goals of the new model?}

\noindent
A: The goal is to \emph{optimize dissemination}, with three sub-goals:
1) make dissemination faster 2) promote open peer review 3) make
dissemination as cheap as the propagation of free speech in society
and 4) guarantee persistence and availability of content over time.

We acknowledge that previous work already exists to address \#1 and
\#2, however we believe that our proposal is the first that proposes a
concrete framework that addresses all 4 goals coherently.

\-

\noindent
\textbf{Q: What exactly are the ``negative incentives to publication'' (Section
6) that are addressed in the paper?}

\noindent
A: We really mean ``incentives to not publish.'' We are focusing here
on the publisher-created culture which only acknowledges the existence
of scientific works once they appear on the publisher's document
listings. Because of this culture, authors refrain from publishing (=
making their work publicly available) before the publisher-mandated
publication process completes, because otherwise other unscrupulous
authors could reuse the work, advance their research faster and
publish the results under their own name. This problem is avoided by
providing near-instantaneous secure authentication of document
existence, which authors can use immediately after completing their
work and before they receive reviews from peers.

\-

\noindent
\textbf{Q: The ACM digital library already support comments on papers. Why is
this not used more?}

\noindent
A: Because such comments are ``held'' by the ACM, who can decide to
remove/modify/hide them arbitrarily at any time.  These comments also
do not increase the visibility and reputation of the reviewer. We
propose to treat reviews as publications, which generates an incentive
to write reviews in the first place.

\-

\noindent
\textbf{Q: who is in a position to spots conflicts of interests between
anonymized publications and anonymized reviewers?}

\noindent
A: Who is in this position now? The academic publishers have already given
the responsibility to spot and resolve conflicts to the program chairs
and/or journal editors; esteemed members of the research community
that are in charge now and would remain so in our model.

\-

\noindent
\textbf{Q:} “we suggest that universities propose \emph{review
  anonymization} as an optional service to their local research
staff. Using this service, the reviews are public but published under
a pseudonym, and the library is responsible for keeping track of real
identities. This escrow service for researcher identities would also
protect accountability: consistently poor reviews could then be
tracked to their real authors after suitably authorized
investigations.” \textbf{Authorized by whom?}

\noindent
See \cref{sec:anon}.

\-

\noindent
\textbf{Q: How does the proposed model compare to similar, previous proposals
for open peer review?}

\noindent
A: Previous proposals for open peer review have focused on the quality
of research evaluation, and the transparency of the review
process. Our proposal embraces these goals, but focuses on a different
aspect: cheap dissemination and persistence of content over time. Our
proposal and previous work on open peer review are thus complementary
to each other (\cf \cref{sec:rel}).

\-

\noindent
\textbf{Q: how to ensure the quality of the formatting, typography, language
etc?}

\noindent
A: To answer this we must consider separately language, and
formatting/typography.
For formatting/typography, we expect each community to settle on a set
of standard document templates. Again we capitalize on a recent
development: since the turn of the century, most authors have learned
to use templates for formatting and typography and are able to produce
print-ready PDF documents of their manuscripts. This is particularly
true of mathematics, logic and computer science, but our experience
shows the same results can be obtained in most other scientific
fields.

For language, a large majority of articles are typically corrected
prior to submission already; most conference articles for instance are
of sufficient linguistic quality despite the lack of external
editing. When in doubt, or in regions where documents are published in
a language where most local practitioners are not proficient, a free
market of proofreading services would develop, in extension to the
services already available for publishing academic theses. These
services are typically much cheaper per article than the price charged
by publishers for publication/access.

\-

\noindent
\textbf{Q: What would motivate a researcher to do a review of another work?}

\noindent
A: Foremost reputation and visibility, but also the intention to
inform a fellow researcher about the quality/correctness and usefulness
of a paper. As reviews are considered publications, writing a review
can increase one's reputation.

\-

\noindent
\textbf{Q: Who would 'force' the researcher to actually do a review?}

\noindent
A: Researchers are currently not ``forced'' to
do reviews by journals or conference organizers, and our proposed
model does not force reviewers either. However the transparency of
reviews will become an incentive to participate. See the previous question.

\-

\noindent
\textbf{Q: What could be a minimum viable prototype of this publication model
to get people started?}

\noindent
A: We have answered this already partially in the TRUST article: the
minimum is a working certificate-of-existence authenticator, a lookup
service from document handles to document contents, and one or more
query engines. We estimate that the effort to reach a prototype is
less than a man-year effort by a computer engineering graduate with a
specialization in software engineering and distributed systems.

\-

\noindent
\textbf{Q: Does Academia 2.0 require changes to the current promotion and
tenure system?  What are those changes? Are such changes feasible?}

\noindent
A: We do not foresee any particular changes. Our proposal implies that
communities of reviewers will form, with their own saved queries (all
papers reviewed by members of our community) and viewing
portals (\cf \eg \cref{sec:newgen}). Consensus will emerge as to which communities are
particularly good at spotting relevant work in their field, and
membership to these communities can then be used for tenure
evaluation.

\-

\noindent
\textbf{Q: How long does it take for the ``cream to rise to the top''?}

\noindent
A: Assuming this question relates to the duration between the point in
time where a document becomes visible on the network, and the point
where it starts to appear in search results: it will depend on the
propagation of document copies between data stores, and the synchronization
rate of document caches throughout the network. Once a document
becomes accessible by a query engine, it becomes immediately candidate
for search results; the time for it to reach the ``top'' of a search
result list is then dependent on the rate of publication of positive
reviews. Assuming an author of a newly published work does due
diligence and announces their work to their community of peers through
conventional channels (mailing lists, public announcements), reviews
could start flowing in within a few days to a few weeks.

\-

\noindent
Q: About preservation of content: \textbf{Relying on author pages seems dicey. I am
also not convinced that all institutions will be so concerned about
preservation of material (sure, major libraries at major research
institution will, but what about the lone researcher at a small
teaching university?)}

\noindent
A: Data stores would accept external submissions, in the same way as
already done for Cornell University's arXiv service. See also \cref{sec:storage}.

\-

\noindent
\textbf{Q: How would double blind reviews work?}

\noindent
A: Double blind reviews would be organized by self-appointed
committees, with a similar process as already used today in
conferences and existing journals. See \cref{sec:dbr}.

\-

\noindent
\textbf{Q: a hallmark of the journal review process is the mediated
interaction between authors and reviewers.  How does this work in
Academia 2.0?}

\noindent
A: This is organized by workflows, in similar ways as currently
organized. See \cref{sec:reviews}.

\-

\noindent
\textbf{Q: I disagree that computer engineering is the scientific field that
has the most potential to try out and evaluate the model. I think
fields such as psychology, sociology, and physics are more likely to
try out the model.}

\noindent
A: During an initial development phase, the tools may not be very
user-friendly. In our view, practitioners in computer engineering will
be more accepting of imperfections and more willing to contribute
improvements the (open source) tools as needed. Of course, it may well
be that other fields could also \emph{use} the initial tools early
simultaneously with computer engineering.

\-

\noindent
\textbf{Q: should reviewer comments be published in full?}

\noindent
A: This largely
  depends on what queries the scientific community will deem useful
  to search, filter, rank and qualify works in query engines.

  The key question is: what are the meaningful fields of a review
  object that are relevant to searching, filtering, ranking and
  qualifying scientific work long after the work was published and the
  review was authored?

  With only numerical ``evaluation grades'', the meaning of these
  grades may be forgotten over time, lest a semantic model is provided
  alongside with the reviews to interpret the grades. Worse, grades do
  not convey the subtleties often intended by reviewers. By publishing
  ``plain text'' comments from reviewers, and assuming computers get
  better in analyzing semantics from human languages, richer and more
  meaningful query engines can be developed.

\section*{Acknowledgements}
\addcontentsline{toc}{section}{Acknowledgements}

The authors are grateful to the anonymous reviewers
of the ACM 2014 TRUST workshop for their thoughtful
and extensive suggestions.

\printindex
\addcontentsline{toc}{section}{Index}

\newcommand{\etalchar}[1]{#1} 
\addcontentsline{toc}{section}{References}
\bibliographystyle{is-plainurl}
\bibliography{doc}

\begin{thebibliography}{10}
\ifx \showCODEN  \undefined \def \showCODEN #1{CODEN #1}  \fi
\ifx \showISBN   \undefined \def \showISBN  #1{ISBN #1}   \fi
\ifx \showISSN   \undefined \def \showISSN  #1{ISSN #1}   \fi
\ifx \showLCCN   \undefined \def \showLCCN  #1{LCCN #1}   \fi
\ifx \showPRICE  \undefined \def \showPRICE #1{#1}        \fi
\ifx \showURL    \undefined \def \showURL {URL }          \fi
\ifx \path       \undefined \input path.sty               \fi
\ifx \ifshowURL \undefined
     \newif \ifshowURL
     \showURLtrue
\fi

\bibitem{bachmann.11.fcn}
Talis Bachmann.
\newblock Fair open evaluation may need some temporarily hidden authorship,
  caution about counting the votes, and transparency of the full
  pre-publication paperwork.
\newblock {\em Frontiers in Computational Neuroscience}, 5\penalty0 (61), 2011.
\newblock \showISSN{1662-5188}.
\newblock \href {http://dx.doi.org/10.3389/fncom.2011.00061}
  {\path{doi:10.3389/fncom.2011.00061}}.

\bibitem{birukou.11.fcn}
Aliaksandr Birukou, Joseph~Rushton Wakeling, Claudio Bartolini, Fabio Casati,
  Maurizio Marchese, Katsiaryna Mirylenka, Nardine Osman, Azzurra Ragone,
  Carles Sierra, and Aalam Wassef.
\newblock Alternatives to peer review: novel approaches for research
  evaluation.
\newblock {\em Frontiers in Computational Neuroscience}, 5\penalty0 (56), 2011.
\newblock \showISSN{1662-5188}.
\newblock \href {http://dx.doi.org/10.3389/fncom.2011.00056}
  {\path{doi:10.3389/fncom.2011.00056}}.

\bibitem{blank.91}
Rebecca~M. Blank.
\newblock The effects of double-blind versus single-blind reviewing:
  Experimental evidence from the american economic review.
\newblock {\em The American Economic Review}, 81\penalty0 (5):\penalty0
  1041--1067, December 1991.

\bibitem{blibech.06}
Kaouthar Blibech and Alban Gabillon.
\newblock A new timestamping scheme based on skip lists.
\newblock In Marina Gavrilova, Osvaldo Gervasi, Vipin Kumar, C.J.Kenneth Tan,
  David Taniar, Antonio Lagan, Youngsong Mun, and Hyunseung Choo, editors, {\em
  Computational Science and Its Applications - ICCSA 2006}, volume 3982 of {\em
  Lecture Notes in Computer Science}, pages 395--405. Springer Berlin
  Heidelberg, 2006.
\newblock \showISBN{978-3-540-34075-1}.
\newblock \href {http://dx.doi.org/10.1007/11751595_43}
  {\path{doi:10.1007/11751595_43}}.

\bibitem{buchmann.09}
Johannes Buchmann, Erik Dahmen, and Michael Szydlo.
\newblock Hash-based digital signature schemes.
\newblock In DanielJ. Bernstein, Johannes Buchmann, and Erik Dahmen, editors,
  {\em Post-Quantum Cryptography}, pages 35--93. Springer Berlin Heidelberg,
  2009.
\newblock \showISBN{978-3-540-88701-0}.
\newblock \href {http://dx.doi.org/10.1007/978-3-540-88702-7_3}
  {\path{doi:10.1007/978-3-540-88702-7_3}}.

\bibitem{devries.09}
Dennis~R. De~Vries, Elizabeth~A. Marschall, and Roy~A. Stein.
\newblock Exploring the peer review process: What is it, does it work, and can
  it be improved?
\newblock {\em Fisheries}, 34\penalty0 (6):\penalty0 270--279, 2009.
\newblock \href {http://dx.doi.org/10.1577/1548-8446-34.6.270}
  {\path{doi:10.1577/1548-8446-34.6.270}}.

\bibitem{florian.12.fcn}
Razvan~Valentin Florian.
\newblock Aggregating post-publication peer reviews and ratings.
\newblock {\em Frontiers in Computational Neuroscience}, 6\penalty0 (31), 2012.
\newblock \showISSN{1662-5188}.
\newblock \href {http://dx.doi.org/10.3389/fncom.2012.00031}
  {\path{doi:10.3389/fncom.2012.00031}}.

\bibitem{ghosh.12.fcn}
Satrajit~S Ghosh, Arno Klein, Brian Avants, and K.~Jarrod Millman.
\newblock Learning from open source software projects to improve scientific
  review.
\newblock {\em Frontiers in Computational Neuroscience}, 6\penalty0 (18), 2012.
\newblock \showISSN{1662-5188}.
\newblock \href {http://dx.doi.org/10.3389/fncom.2012.00018}
  {\path{doi:10.3389/fncom.2012.00018}}.

\bibitem{haber.91}
Stuart Haber and W.~Scott Stornetta.
\newblock How to time-stamp a digital document.
\newblock {\em Journal of Cryptology}, 3\penalty0 (2):\penalty0 99--111, 1991.
\newblock \showISSN{1432-1378}.
\newblock \href {http://dx.doi.org/10.1007/BF00196791}
  {\path{doi:10.1007/BF00196791}}.

\bibitem{hartshorne.12.fcn}
Joshua Hartshorne and Adena Schachner.
\newblock Tracking replicability as a method of post-publication open
  evaluation.
\newblock {\em Frontiers in Computational Neuroscience}, 6\penalty0 (8), 2012.
\newblock \showISSN{1662-5188}.
\newblock \href {http://dx.doi.org/10.3389/fncom.2012.00008}
  {\path{doi:10.3389/fncom.2012.00008}}.

\bibitem{hunter.12.fcn}
Jane Hunter.
\newblock Post-publication peer review: Opening up scientific conversation.
\newblock {\em Frontiers in Computational Neuroscience}, 6\penalty0 (63), 2012.
\newblock \showISSN{1662-5188}.
\newblock \href {http://dx.doi.org/10.3389/fncom.2012.00063}
  {\path{doi:10.3389/fncom.2012.00063}}.

\bibitem{iettogillies.12.fcn}
Grazia Ietto-Gillies.
\newblock The evaluation of research papers in the xxi century. the open peer
  discussion system of the world economics association.
\newblock {\em Frontiers in Computational Neuroscience}, 6\penalty0 (54), 2012.
\newblock \showISSN{1662-5188}.
\newblock \href {http://dx.doi.org/10.3389/fncom.2012.00054}
  {\path{doi:10.3389/fncom.2012.00054}}.

\bibitem{kravitz.11.fcn}
Dwight Kravitz and Chris~I Baker.
\newblock Toward a new model of scientific publishing: Discussion and a
  proposal.
\newblock {\em Frontiers in Computational Neuroscience}, 5\penalty0 (55), 2011.
\newblock \showISSN{1662-5188}.
\newblock \href {http://dx.doi.org/10.3389/fncom.2011.00055}
  {\path{doi:10.3389/fncom.2011.00055}}.

\bibitem{kreiman.11.fcn}
Gabriel Kreiman and John Maunsell.
\newblock Nine criteria for a measure of scientific output.
\newblock {\em Frontiers in Computational Neuroscience}, 5\penalty0 (48), 2011.
\newblock \showISSN{1662-5188}.
\newblock \href {http://dx.doi.org/10.3389/fncom.2011.00048}
  {\path{doi:10.3389/fncom.2011.00048}}.

\bibitem{kriegeskorte.12.fcn2}
Nikolaus Kriegeskorte.
\newblock Open evaluation ({OE}): A vision for entirely transparent
  post-publication peer review and rating for science.
\newblock {\em Frontiers in Computational Neuroscience}, 6\penalty0 (79), 2012.
\newblock \showISSN{1662-5188}.
\newblock \href {http://dx.doi.org/10.3389/fncom.2012.00079}
  {\path{doi:10.3389/fncom.2012.00079}}.

\bibitem{kriegeskorte.12.fcn}
Nikolaus Kriegeskorte, Alexander Walther, and Diana Deca.
\newblock An emerging consensus for open evaluation: 18 visions for the future
  of scientific publishing.
\newblock {\em Frontiers in Computational Neuroscience}, 6\penalty0 (94), 2012.
\newblock \showISSN{1662-5188}.
\newblock \href {http://dx.doi.org/10.3389/fncom.2012.00094}
  {\path{doi:10.3389/fncom.2012.00094}}.

\bibitem{lee.12.fcn}
Christopher Lee.
\newblock Open peer review by a selected-papers network.
\newblock {\em Frontiers in Computational Neuroscience}, 6\penalty0 (1), 2012.
\newblock \showISSN{1662-5188}.
\newblock \href {http://dx.doi.org/10.3389/fncom.2012.00001}
  {\path{doi:10.3389/fncom.2012.00001}}.

\bibitem{lemire.11.dbpr}
Daniel Lemire.
\newblock The case against double blind peer review [online].
\newblock April 2011.

\bibitem{poeschl.12.fcn}
Ulrich P{\"o}schl.
\newblock Multi-stage open peer review: scientific evaluation integrating the
  strengths of traditional peer review with the virtues of transparency and
  self-regulation.
\newblock {\em Frontiers in Computational Neuroscience}, 6\penalty0 (33), 2012.
\newblock \showISSN{1662-5188}.
\newblock \href {http://dx.doi.org/10.3389/fncom.2012.00033}
  {\path{doi:10.3389/fncom.2012.00033}}.

\bibitem{poss.14.trust}
Raphael Poss, Sebastian Altmeyer, Mark Thompson, and Rob Jelier.
\newblock {Academia 2.0}: removing the publisher middle-man while retaining
  impact.
\newblock In {\em Proc 1st ACM SIGPLAN Workshop on Reproducible Research
  Methodologies and New Publication Models in Computer Engineering (TRUST'14)}.
  ACM, Edinburgh, UK, June 2014.
\newblock \showISBN{978-1-4503-2951-4}.
\newblock \href {http://dx.doi.org/10.1145/2618137.2618139}
  {\path{doi:10.1145/2618137.2618139}}.

\bibitem{poss.14.ads}
{Raphael~`kena'} Poss.
\newblock Aca2: protocol specification (draft), May 2014.
\newblock Available from:
  \url{http://science.raphael.poss.name/aca2-draft-spec.html}.

\bibitem{priem.12.fcn}
Jason Priem and Bradley~H. Hemminger.
\newblock Decoupling the scholarly journal.
\newblock {\em Frontiers in Computational Neuroscience}, 6\penalty0 (19), 2012.
\newblock \showISSN{1662-5188}.
\newblock \href {http://dx.doi.org/10.3389/fncom.2012.00019}
  {\path{doi:10.3389/fncom.2012.00019}}.

\bibitem{sandewall.12.fcn}
Erik Sandewall.
\newblock Maintaining live discussion in two-stage open peer review.
\newblock {\em Frontiers in Computational Neuroscience}, 6\penalty0 (9), 2012.
\newblock \showISSN{1662-5188}.
\newblock \href {http://dx.doi.org/10.3389/fncom.2012.00009}
  {\path{doi:10.3389/fncom.2012.00009}}.

\bibitem{walther.12.fcn}
Alexander Walther and Jasper Jacobus~Franciscus van~den Bosch.
\newblock {FOSE}: A framework for open science evaluation.
\newblock {\em Frontiers in Computational Neuroscience}, 6\penalty0 (32), 2012.
\newblock \showISSN{1662-5188}.
\newblock \href {http://dx.doi.org/10.3389/fncom.2012.00032}
  {\path{doi:10.3389/fncom.2012.00032}}.

\bibitem{wicherts.12.fcn}
Jelte~M Wicherts, Rogier~A Kievit, Marjan Bakker, and Denny Borsboom.
\newblock Letting the daylight in: Reviewing the reviewers and other ways to
  maximize transparency in science.
\newblock {\em Frontiers in Computational Neuroscience}, 6\penalty0 (20), 2012.
\newblock \showISSN{1662-5188}.
\newblock \href {http://dx.doi.org/10.3389/fncom.2012.00020}
  {\path{doi:10.3389/fncom.2012.00020}}.

\bibitem{wilson.12}
Mark~C. Wilson.
\newblock Filling a much-needed gap [online].
\newblock January 2012.

\bibitem{zimmermann.12.fcn}
Jan Zimmermann, Alard Roebroeck, Kamil Uludag, Alexander~T Sack, Elia
  Formisano, Bernadette Jansma, Peter De~Weerd, and Rainer Goebel.
\newblock Network-based statistics for a community driven transparent
  publication process.
\newblock {\em Frontiers in Computational Neuroscience}, 6\penalty0 (11), 2012.
\newblock \showISSN{1662-5188}.
\newblock \href {http://dx.doi.org/10.3389/fncom.2012.00011}
  {\path{doi:10.3389/fncom.2012.00011}}.

\end{thebibliography}

\end{document}